\documentclass[aps,pre,amsmath,amssymb,reprint,onecolumn,floatfix,longbibliography,notitlepage]{revtex4-1}

\usepackage{bm}
\usepackage{graphicx}

\usepackage{color}

\begin{document}

\title{
Energy-based analysis and anisotropic spectral distribution of internal gravity waves in strongly stratified turbulence
}

\author{Naoto Yokoyama}
\email{yokoyama@me.es.osaka-u.ac.jp}
\affiliation{Department of Mechanical Science and Bioengineering, Osaka University, Toyonaka 560-8531, Japan}

\author{Masanori Takaoka}
\email{mtakaoka@mail.doshisha.ac.jp}
\affiliation{Department of Mechanical Engineering, Doshisha University, Kyotanabe 610-0394, Japan}

\date{\today}

\begin{abstract}
Stratified turbulence shows
scale- and direction-dependent anisotropy
and the coexistence of weak turbulence of internal gravity waves and strong turbulence of eddies.
Straightforward application of standard analyses developed in isotropic turbulence
sometimes masks important aspects of the anisotropic turbulence.
To capture detailed structures of the energy distribution in the wave-number space,
 it is indispensable to examine the energy distribution
 with non-integrated spectra by fixing the codimensional wave-number component
 or
 in the two-dimensional domain spanned by both the horizontal and vertical wave numbers.
Indices which separate the range of the anisotropic weak-wave turbulence in the wave-number space
 are proposed based on the decomposed energies.
In addition,
the dominance of the waves in the range is also verified
by the small frequency deviation from the linear dispersion relation.
In the wave-dominant range,
 the linear wave periods given by the linear dispersion relation are smaller
 than approximately one third of the eddy-turnover time.
The linear wave periods reflect the anisotropy of the system,
while the isotropic Brunt-V\"ais\"al\"a period is used to evaluate the Ozmidov wave number,
which is necessarily isotropic.
It is found that
the time scales in consideration of the anisotropy of the flow field
must be appropriately selected
to obtain the critical wave number separating the weak-wave turbulence.
\end{abstract}

\maketitle

\section{Introduction}

Turbulence in nature essentially has anisotropy
especially in large scales.
Theoretical approaches in turbulence researches
originate from Kolmogorov's local isotropy hypothesis,
and have been extended to researches in anisotropic turbulence systems.
Numerical simulations of high Reynolds-number turbulent flows and their analyses
are developed also in homogeneous statistically-isotropic turbulence systems,
and are often incorporated in anisotropic turbulence systems simply.
It is essential to introduce appropriate analytical tools
which do not diminish scale- and direction-dependent anisotropic properties
in the anisotropic turbulence.

Stratified turbulence is one of the most fundamental turbulence systems which have statistical anisotropy,
and is observed in the oceans and the atmosphere.
The gravity produces the density or thermal stratification,
and makes statistical differences in energy distribution
between the vertical direction and the horizontal direction.
The breaking of the internal gravity waves affects the global climate and our lives;
the upwelling due to breaking is an important part of the thermohaline circulation in the oceans~\cite{MUNK1966707},
and the breaking in the atmosphere sometimes causes clear-air turbulence
that may expose aircraft flight to risk~\cite{doi:10.1175/1520-0469(2000)057<1105:OOADCA>2.0.CO;2}.
The breaking corresponds the energy transfer from waves to vortices.

Various kinds of energy spectra have been reported in observations,
experiments, and simulations of stratified turbulence.
The variety is derived from
the physical mechanisms, the length scales, and other parameters.
Such different energy spectra can coexist,
and the coexistence
is obtained in atmospheric observations and numerical simulations~\cite{nastrom1984kinetic,FLM:8539071,FLM:409533}.
For example,
a kinetic-energy spectrum observed in atmospheric flows
has a power law
$K_{\perp}(k_{\perp}) \propto  k_{\perp}^{-3}$
at large scales~\cite{nastrom1984kinetic}.
Here, $k_{\perp}$ is a horizontal wave number,
and $K_{\perp}(k_{\perp})$ is the horizontal kinetic-energy spectrum
as a function of $k_{\perp}$.
Another power law
$K_{\perp}(k_{\perp}) \propto k_{\perp}^{-5/3}$
is also observed at mesoscales,
and the same power law is obtained analytically and numerically~\cite{FLM:409533}.
Observation and theoretical prediction also have a variety of the kinetic-energy spectrum as a function of the vertical wave number $k_{\|}$:
the breaking of the internal gravity waves makes the total kinetic-energy spectrum
$K(k_{\|}) \propto k_{\|}^{-3}$~\cite{doi:10.1175/1520-0469(1987)044<1404:EFASSO>2.0.CO;2}
for example.
The Bolgiano-Obukhov phenomenology predicts coexistence of two power-laws
in kinetic spectra: $K(k)\propto k^{-11/5}$ for $k<k_{\mathrm{B}}$
and $K(k)\propto k^{-5/3}$ for $k>k_{\mathrm{B}}$,
where $k_{\mathrm{B}}$ is the Bolgiano wave number~\cite{JGR:JGR897,Obukhov1959}.
The pioneering work for the two-dimensional energy spectrum
of the internal gravity waves observed in the ocean
is the Garrett-Munk spectrum,
which has $K(k_{\perp}, k_{\|}) \propto k_{\perp}^{-2} k_{\|}^{-1}$ at relatively large wave numbers~\cite{GM_ARF}.
The weak turbulence theory predicts a variety of power laws including the Garrett-Munk spectrum~\cite{iwthLPTN}.
A spectral model that allows even variability was proposed~\cite{JGRD:JGRD2467}.
In this way,
the kinetic-energy spectra as well as the potential-energy spectra
are diverse,
and the diversity may result from the boundary conditions
and the magnitude relation between the horizontal wave number and the vertical wave number.
On the other hand,
when the stratification is relatively weak,
the vortices are dominant in the flow,
 and the three-dimensional isotropic Kolmogorov turbulence appears.
Then, the energy spectrum
shows the Kolmogorov's power law $K(k) \propto k^{-5/3}$.

To elucidate the variability of the energy spectra at the small wave numbers
and to consistently observe them,
the dominant physical mechanism at a wave number is required to be evaluated.
In this case,
the one-dimensionalized energy spectra
such as $K_{\perp}(k_{\perp})$ obtained by integration over $k_{\|}$
cannot properly reflect
the energy distribution in the anisotropic turbulence.
The wave-number range where one of the physical mechanisms framing the anisotropic turbulence is dominant
should be identified in the $k_{\perp}$--$k_{\|}$ space.

It is the general practice to focus on time scales
to find a dominant mechanism in complex dynamical systems
which have multiple physics~\cite{kevorkian2012multiple}.
In the three-dimensional isotropic Kolmogorov turbulence,
for example,
the eddies in the inertial subrange have
the eddy-turnover time shorter than the dissipation time,
while the dissipation time is shorter than the eddy-turnover time in the dissipation range.
The Kolmogorov wave number,
which separates the inertial subrange and the dissipation range,
is defined
so that the eddy-turnover time is equal to the dissipation time.

The weak turbulence theory,
which has been successfully applied
to the statistical description of the nonlinear energy transfers
among weakly-coupled dispersive waves,
assumes that
the linear time scale evaluated by the linear dispersion relation
is much smaller than the nonlinear time scale of energy transfers.
However,
the linear time scale becomes comparable with the nonlinear time scale,
and the assumption of the weak nonlinearity is violated
either at small or large wave numbers
in most of the wave turbulence systems~\cite{Biven200128,newell01,Biven200398}.
As a result,
the weak-wave turbulence and the strong turbulence
coexist in many wave turbulence systems
such as stratified turbulence considered here,
rotating turbulence~\cite{PhysRevFluids.2.092602}, magnetohydrodynamic turbulence~\cite{PhysRevX.8.031066}, elastic-wave turbulence~\cite{PhysRevE.89.012909} and quantum turbulence~\cite{Vinen2002}.

In stratified turbulence,
the Brunt-V\"ais\"al\"a period and the eddy-turnover time
have respectively been used as the linear and nonlinear time scales.
The Ozmidov wave number defined as the wave number
at which these two time scales are comparable
has been considered as the critical wave number
that separates the strongly anisotropic turbulence and the isotropic Kolmogorov turbulence~\cite{ozmidovscale}.
In fact,
the wave numbers much larger than the Ozmidov wave number,
the stratification can be almost negligible,
and the isotropic Kolmogorov turbulence appears.
The buoyancy wave number,
which is defined by the characteristic horizontal velocity and the Brunt-V\"ais\"al\"a frequency,
gives the scale of the shear layers and the breaking of the internal gravity waves~\cite{doi:10.1063/1.3599699}.

On the other hand,
the weak-wave turbulence does not appear at all the wave numbers smaller than the Ozmidov wave number or the buoyancy wave number.
The anisotropic quasi-two-dimensional turbulence
such as
the layer-wise two-dimensional turbulence~\cite{doi:10.1175/1520-0469(1983)040<0749:STATMV>2.0.CO;2} and the pancake turbulence~\cite{:/content/aip/journal/pof2/13/6/10.1063/1.1369125}
exists at such small wave numbers.
Neither the Ozmidov wave number nor the buoyancy wave number can identify the wave-number range
where statistically-anisotropic gravity-wave turbulence is dominant
because of the isotropy assumed in their derivations.
The anisotropy of the time scales can be introduced
by using the period given by the linear dispersion relation
 instead of the Brunt-V\"ais\"al\"a period
 as the linear time scale~\cite{nazarenko3488critical}.
 The wave number
 at which the period given by the linear dispersion relation
and the eddy-turnover time are comparable
can separate the weak-wave turbulence and the isotropic or anisotropic strong turbulence
in magnetohydrodynamic turbulence~\cite{PhysRevX.8.031066,goldreich1995toward,PhysRevLett.110.145002,PhysRevLett.116.105002}.
However, it is not clear in rotating turbulence~\cite{:/content/aip/journal/pof2/26/3/10.1063/1.4868280}.

In this paper,
direct numerical simulations of strongly stratified turbulence are performed,
and anisotropic properties of internal gravity-wave turbulence
are characterized by distribution and decomposition of energy.
The organization of the paper is as follows.
The numerical scheme of the direct numerical simulations
and decomposition of the wave-number space and the flow field are shown in Sec.~\ref{sec:formulation},
where some definitions of the energies to characterize the anisotropic weak-wave turbulence are provided.
The numerical results are exhibited in Sec.~\ref{sec:numericalresults}.
Indices to identify the range of the anisotropic internal gravity-wave turbulence
are proposed,
and the range is examined in the two-dimensional domain spanned by both of the horizontal and vertical wave numbers.
The last section is devoted to the summary.

\section{Formulation}
\label{sec:formulation}

\subsection{Numerical scheme}
\label{ssec:scheme}

 Incompressible flows in stably stratified background flow in the $z$ direction is considered.
 Under the Boussinesq approximation,
 the governing equation for the velocity $\bm{u}$ and buoyancy $b$
 is given as follows:
 \begin{subequations}
  \begin{align}
   &
   \frac{\partial}{\partial t} \bm{u} + (\bm{u} \cdot \nabla) \bm{u}
   = -\nabla p + b \bm{e}_z + \nu \nabla^2 \bm{u}
   + \bm{f}
   ,
   \\
   &
   \nabla \cdot \bm{u} = 0
   ,
   \\
   &
   \frac{\partial}{\partial t} b + (\bm{u} \cdot \nabla) b
   = - N^2 \bm{u} \cdot \bm{e}_z + \kappa \nabla^2 b
   .
  \end{align}%
  \label{eq:gov}%
 \end{subequations}%
The buoyancy $b$ is given as
 $b = -g \theta^{\prime} / \theta_0$
 in atmospheric flows, for example,
 where $g$, $\theta^{\prime}$, and $\theta_0$ are respectively
 the gravity acceleration, the temperature fluctuation, and the mean temperature.
The Brunt-V\"ais\"al\"a frequency $N$ is assumed to be constant.
The external force $\bm{f}$ is added
to obtain the non-equilibrium statistically-steady state.
The kinematic viscosity and the diffusion constant
 are respectively denoted by $\nu$ and $\kappa$.

 In this work,
 direct numerical simulations of Eq.~(\ref{eq:gov})
 are performed in a periodic box with the side $2\pi$.
 The Fourier coefficients of the dependent variables appearing in Eq.~\eqref{eq:gov},
 $\widetilde{\bm{u}}_{\bm{k}}$, $\widetilde{p}_{\bm{k}}$, and $\widetilde{b}_{\bm{k}}$,
 are used,
 and the tildes are omitted below.
 The pseudo-spectral method with aliasing removal due to the phase shift
 is employed to evaluate the nonlinear term.
The Runge-Kutta-Gill method is used for the time integration.
The external force is added in the wave-number space
to the wave-number mode in $k_{\mathrm{f}}-1/2 \leq |\bm{k}| < k_{\mathrm{f}}+1/2$,
where the forced wave number $k_{\mathrm{f}}$ is set to $4$.
The external force is generated by the Ornstein-Uhlenbeck process~\cite{FLM:8539071}
as follows.
The colored noise $\hat{\bm{f}}_{\bm{k}} = (\hat{f}_{x \bm{k}}, \hat{f}_{y \bm{k}}, 0)$,
which consists of two spatial components each having a complex value,
is obtained
for each wave number
according to the following stochastic differential equation:
 \begin{align}
 \begin{pmatrix}
  d \hat{\bm{f}}_{\bm{k}} \\
  d \hat{\bm{g}}_{\bm{k}}
 \end{pmatrix}
  =
  dt
 \begin{pmatrix}
  -\alpha & 1
  \\
  0 & -\alpha
 \end{pmatrix}
 \begin{pmatrix}
  \hat{\bm{f}}_{\bm{k}} \\
  \hat{\bm{g}}_{\bm{k}}
 \end{pmatrix}
 +
 \begin{pmatrix}
  \bm{0} \\
  \gamma_{\bm{k}} d\bm{W}_{\bm{k}}
 \end{pmatrix}
,
 \end{align}%
where
 $d\bm{W}_{\bm{k}}$ represents the normal random variables with mean $0$ and variance $dt$
 and has four independent components.
 The correlation time of $\hat{\bm{f}}_{\bm{k}}$ is $O(1/\alpha)$,
 and $\alpha$ is set to be $N$ in this paper.
 Because $\langle |\hat{\bm{f}}_{\bm{k}}|^2 \rangle = \gamma_{\bm{k}}^2/\alpha$,
$\gamma_{\bm{k}}$ is used to control the amplitude of the external force.
Finally,
the Fourier coefficient of the external force is set as
$\bm{f}_{\bm{k}} = \hat{\bm{f}}_{\bm{k}} - \bm{k} (\bm{k} \cdot\hat{\bm{f}}_{\bm{k}}) / k^2$
to satisfy the divergence-free condition.

The number of the grid points used is up to $2048^3$.
The low-resolution simulations with $1024^3$ grid points are also used to examine the parameter dependence.
The corresponding largest wave number $k_{\mathrm{max}}$
is approximately $970$ or $480$.
The Brunt-V\"ais\"al\"a frequency is set to $N=10$.
The Prandtl number is set to be unity,
i.e., $\nu=\kappa$,
and $\nu$ is chosen so that $k_{\mathrm{max}} / k_{\eta} \approx 1.2$.
Here, $k_{\eta} = (\overline{\varepsilon}/\nu^3)^{1/4}$ is the Kolmogorov wave number,
and $\overline{\varepsilon}$ denotes the mean dissipation rate of the kinetic energy.
The coefficient $\gamma_{\bm{k}}$ to control the amplitude of the external force
is varied in the simulations with $1024^3$ grids.
The parameters in the numerical simulations
and their definitions which follow those in Ref.~\cite{maffioli_davidson_2016}
are summarized in Table~\ref{tab:parameters}.

\begin{table}[t]
  \caption{Parameters in the numerical simulations.
$Re$: horizontal Reynolds number,
$Re_{\mathrm{b}}$: buoyancy Reynolds number,
$Fr_{\perp}$: horizontal Froude number,
$Fr_{\|}$: vertical Froude number,
$k_{\mathrm{O}}$: Ozmidov wave number,
$k_{\mathrm{b}}$: buoyancy wave number.
The root-mean square of the horizontal velocity is denoted by $u_{\perp \mathrm{rms}}$.
The horizontal and vertical integral length scales, $\ell_{\perp}$ and $\ell_{\|}$, are defined by transverse velocity correlations.
}
  \label{tab:parameters}
  \begin{ruledtabular}
  \begin{tabular}{cccccccc}
   number of grid points & $\gamma_{\bm{k}}$ & $Re$ & $Re_{\mathrm{b}}$ & $Fr_{\perp}$ & $Fr_{\|}$ & $k_{\mathrm{O}}$ & $k_{\mathrm{b}}$ \\
   && $u_{\perp \mathrm{rms}} \ell_{\perp} / \nu$ &
	       $\overline{\varepsilon} / (\nu N^2)$ &
		   $u_{\perp \mathrm{rms}} /(N \ell_{\perp})$ &
		       $u_{\perp \mathrm{rms}} /(N \ell_{\|})$ &
			   $\sqrt{N^3/\overline{\varepsilon}}$ &
			       $N/u_{\perp \mathrm{rms}}$ \\
   \hline
   $2048^3$ & $0.5$ &
   $7.7\times 10^4$ & $2.1$ & $9.0\times 10^{-3}$ & $0.80$ & $490$ & $27$
\\
   $1024^3$ & $0.1$ &
   $1.8 \times 10^4$ & $9.4\times 10^{-2}$ & $4.9 \times 10^{-3}$ & $0.15$ & $2300$ & $76$
\\
   $1024^3$ & $0.2$ &
   $1.1 \times 10^4$ & $0.24$ & $9.4 \times 10^{-3}$ & $0.17$ & $1200$ & $57$
\\
   $1024^3$ & $0.5$ &
   $2.5\times 10^4$ & $0.80$ & $9.6\times 10^{-3}$ & $0.46$ & $500$ & $29$
\\
   $1024^3$ & $1$ &
   $4.1 \times 10^4$ & $1.9$ & $1.4 \times 10^{-2}$ & $1.6$ & $260$ & $15$
\\
   $1024^3$ & $2$ &
   $3.7 \times 10^4$ & $4.7$ & $1.9 \times 10^{-2}$ & $2.0$ & $130$ & $11$
\\
   $1024^3$ & $5$ &
   $2.2 \times 10^4$ & $18$ & $3.5 \times 10^{-2}$ & $1.0$ & $53$ & $8.2$
  \end{tabular}
\end{ruledtabular}
\end{table}

The initial condition of a simulation is
a statistically steady state of the lower-resolution simulation.
Therefore, the small wave-number modes are numerically integrated over a long time as $N t = O(10^3)$.
Because all the simulations relax to statistically steady states
after some times depending on the amplitudes of the external force,
the growth without stationarity reported in Ref.~\cite{smith_waleffe_2002}
was not observed in the simulations.
The time averaging is performed to draw the spectra
for $N t = 100$ with every $12.5$ in the high-resolution simulation.
It might be short to remove the fluctuation at the small wave numbers,
but the results shown in this paper are confirmed to be unchanged in the low-resolution simulations,
where long-time averaging is performed.

\subsection{Ratios of time scales to find the dominant physical mechanism}
\label{ssec:timeratio}

 The Ozmidov wave number $k_{\mathrm{O}}$
 has been considered as a wave number
 which separates the strongly anisotropic range and the isotropic range in the wave-number space.
The Ozmidov wave number is given as a wave number
 at which the Brunt-V\"ais\"al\"a period $1/N$
 and the eddy-turnover time of the three-dimensional (3D) isotropic turbulence $\tau_{\bm{k}} = 1/(ku) = (k^2 \overline{\varepsilon})^{-1/3}$
 are comparable,
i.e.,
 $k_{\mathrm{O}} = \sqrt{N^3/\overline{\varepsilon}}$.
 It should be noted that
 the Ozmidov wave number is independent of the direction of the wave number vector, i.e., isotropic.
 The 3D isotropic Kolmogorov turbulence is expected to dominate
 at the wave numbers larger than $k_{\mathrm{O}}$,
 but $k_{\mathrm{O}}$ does not necessarily determine the wave-number range
 where the weak gravity-wave turbulence is dominant
 because of the lack of the anisotropy.
 The buoyancy wave number $k_{\mathrm{b}} = N/u_{\perp \mathrm{rms}}$
 is another wave number
 that characterizes the transition from the quasi-two-dimensional turbulence to the 3D isotropic turbulence.
 The buoyancy wave number is also isotropic.

 Owing to the anisotropy,
 the spectral structures in the wave-number space should be investigated in the $k_{\perp}$--$k_{\|}$ space.
 The theory of the critical balance
 states that the energy is transferred
 in the transitional wave-number range between the wave-dominant and vortex-dominant ranges~\cite{nazarenko3488critical}.
 In this theory,
 the wave period of the gravity wave given by the linear dispersion relation
 is employed
 as the linear time scale
 instead of the Brunt-V\"ais\"al\"a period.
 Note that the linear dispersion relation is anisotropic.
 Because $k_{\perp} \ll k_{\|}$ and hence $|\bm{u}_{\perp}| \gg |u_{\|}|$ owing to the divergence-free condition were assumed in Refs.~\cite{nazarenko3488critical,nazarenkobook},
 the linear dispersion relation was rewritten
 as $\sigma_{\mathrm{2D}\bm{k}} = Nk_{\perp}/k_{\|}$,
 and the eddy-turnover time of the two-dimensional (2D) turbulence $\tau_{\mathrm{2D}\bm{k}} = 1/(k_{\perp}u_{\perp}) = (k_{\perp}^2 \overline{\varepsilon})^{-1/3}$
 was used as the nonlinear time.
 In the present work,
 since the strong turbulence is not only 2D but also 3D
 and $k_{\perp} \ll k_{\|}$ does not necessarily hold,
 the general linear dispersion relation $\sigma_{\bm{k}} = Nk_{\perp}/k$
 is used to evaluate the linear time.
 Moreover, the eddy-turnover time of the 3D turbulence
 $\tau_{\bm{k}} = (k^2 \overline{\varepsilon})^{-1/3}$
 is used as the nonlinear time.
 Then,
 the nonlinearity is evaluated
 by $\chi_{\bm{k}} = 1/(\sigma_{\bm{k}} \tau_{\bm{k}})$.
 The ratio of the gravity-wave period to the 2D eddy-turnover time
 $\chi_{\mathrm{2D}\bm{k}} = 1/(\sigma_{\mathrm{2D}\bm{k}} \tau_{\mathrm{2D}\bm{k}})$
 is also introduced for reference.

\subsection{Decomposition of turbulent flow}
\label{ssec:decomposition}

  To examine the idea of the critical balance,
  it is indispensable to identify the wave-dominant range.
The Craya-Herring (Cartesian) decomposition and the helical-mode decomposition are used
for the identification
in this paper.

In the Craya-Herring decomposition~\cite{doi:10.1063/1.1694822,FLM:8539071},
an orthonormal basis,
$\bm{e}_1 = \bm{k} \times \bm{e}_z/k_{\perp}$,
$\bm{e}_2 = \bm{k} \times \bm{e}_1 /k$,
and $\bm{e}_3 = \bm{k}/k$,
is introduced.
The two basis vectors
$\bm{e}_1$ and $\bm{e}_2$ are defined
only when $\bm{k}$ and $\bm{e}_z$ are not in parallel,
that is, horizontal component of $\bm{k}$, $\bm{k}_{\perp}$, is non-zero.
The orthogonal basis decomposes the velocity as
 \begin{align}
  \bm{u}_{\bm{k}} =
  \begin{cases}
   u_{\mathrm{v}} \bm{e}_1 +
   u_{\mathrm{w}} \bm{e}_2
   & \text{for } k_{\perp} \neq 0
   \\
   \bm{u}_{\mathrm{s}}
   & \text{for } k_{\perp} = 0
  \end{cases}
.
 \end{align}
The Fourier component of the velocity
is given by two components perpendicular to the wave-number vector $\bm{k}$
because of the incompressibility $\bm{k} \cdot \bm{u}_{\bm{k}} = 0$.
When the wave numbers with $k_{\perp} = 0$ are included,
such decomposition is called the Cartesian decomposition.

When the viscosity and the diffusion are neglected
for $k_{\perp} \neq 0$,
the governing equation~(\ref{eq:gov}) can be linearized as
 \begin{align}
  \frac{\partial u_{\mathrm{v} \bm{k}}}{\partial t} = 0,
  \;
  \frac{\partial u_{\mathrm{w} \bm{k}}}{\partial t} = - \frac{k_{\perp}}{k} b_{\bm{k}},
  \;
  \frac{\partial b_{\bm{k}}}{\partial t} = N^2 \frac{k_{\perp}}{k} u_{\mathrm{w} \bm{k}}
  .
  \label{eq:linearinviscidnondiffusive}%
  \end{align}%
This linear inviscid non-diffusive equation indicates that
$u_{\mathrm{v}} = i \omega_z / k_{\perp}$
is a vortical mode that is not affected by the linear buoyancy term,
and
$u_{\mathrm{w}} = - k u_z / k_{\perp}$
is a wave mode.
Here, $\omega_z$ denotes the $z$ component of the vorticity.
The second equation and the third one in Eq.~(\ref{eq:linearinviscidnondiffusive})
give the linear dispersion relation of the gravity waves: $\sigma_{\bm{k}} = N k_{\perp} / k$.
The velocity for $k_{\perp} = 0$,
$\bm{u}_{\mathrm{s} k_z} = \bm{u}_{\perp}$,
represents a vertically-sheared horizontal flow.
Namely, the Cartesian decomposition simply represents
the decomposition of the velocity into the vortices, the waves and the shear flows
at the lowest order.
The Cartesian decomposition is equivalent to the normal-mode decomposition~\cite{waite_bartello_2006}.

The helical-mode decomposition
has also been used for the decomposition of the velocity.
In the helical-mode decomposition,
the basis $\bm{h}_{\pm} = (\bm{e}_2 \mp i \bm{e}_1)/\sqrt{2}$
is the eigen vector for the curl operation,
$i \bm{k} \times \bm{h}_{\pm} = \pm k \bm{h}_{\pm}$.
Then,
the velocity is decomposed as $\bm{u} = \xi_+ \bm{h}_+ + \xi_- \bm{h}_-$,
where
 $\xi_{\pm} = \bm{u} \cdot \bm{h}_{\mp}$
 is the helical-mode intensity.
Note that $\bm{h}_{\pm} \cdot \bm{h}_{\pm} = 0$ and $\bm{h}_{\pm} \cdot \bm{h}_{\mp} = 1$.

 A wave-number mode $\bm{k}$ has a total energy $E_{\bm{k}}$,
 which is sum of the kinetic energy
$K_{\bm{k}} = \langle|\bm{u}_{\bm{k}}|^2\rangle/2$
and potential energy
$V_{\bm{k}} = \langle|b_{\bm{k}}|^2\rangle/(2N^2)$.
The kinetic energy can be given by
horizontal kinetic energy
$K_{\perp \bm{k}} = K_{x \bm{k}} + K_{y \bm{k}} =(\langle|u_{x \bm{k}}|^2\rangle + \langle|u_{y \bm{k}}|^2\rangle)/2$
and vertical kinetic energy
$K_{\| \bm{k}} = K_{z \bm{k}} = \langle|u_{z \bm{k}}|^2\rangle/2$,
focused on the direction of the velocity.
Similarly,
the Cartesian decomposition defines
vortical kinetic energy
 $K_{\mathrm{v} \bm{k}} = \langle|u_{\mathrm{v} \bm{k}}|^2\rangle/2
 = \langle|\omega_{z \bm{k}}|^2\rangle/(2 k_{\perp}^2)$,
wave kinetic energy
 $K_{\mathrm{w} \bm{k}} = \langle|u_{\mathrm{w} \bm{k}}|^2\rangle/2
 = k^2 \langle|u_{z \bm{k}}|^2\rangle/(2 k_{\perp}^2)$,
and
shear kinetic energy
 $K_{\mathrm{s} k_z} = \langle|\bm{u}_{\mathrm{s} k_z}|^2\rangle/2 = \langle|\bm{u}_{\perp k_z}|^2\rangle/2$.
Because the shear flow is defined only for $k_{\perp} = 0$,
it depends only on $k_z$.
Moreover,
according to the helical-mode decomposition,
the kinetic energy in the $m$ direction,
where $m=x,y,z$,
can be written as
 \begin{align}
  K_{m \bm{k}}
  &= \frac{K(k)}{8\pi k^2} \left(1- \frac{k_{m}^2}{k^2}\right)
  + \frac{1}{2} \left(K_{\bm{k}} - \frac{K(k)}{4\pi k^2}\right) \left(1- \frac{k_{m}^2}{k^2}\right)
  + \mathrm{Re} [Z_{\bm{k}} h_{+ m \bm{k}}^2]
  .
  \label{eq:Kalphak}
 \end{align}
Here,
$K(k)$ is the one-dimensionalized energy spectrum,
and $Z_{\bm{k}} = \langle\xi_{+ \bm{k}} \xi_{- \bm{k}}^{\ast} \rangle
 = K_{\mathrm{w} \bm{k}} - K_{\mathrm{v} \bm{k}} + i \mathrm{Re} \langle u_{\mathrm{v} \bm{k}}^{\ast} u_{\mathrm{w} \bm{k}}\rangle$.
The terms in the right-hand side of Eq.~(\ref{eq:Kalphak})
represent
isotropic part,
directional anisotropic part with respect to the direction of $\bm{k}$,
and polarization anisotropic part with respect to the direction of $\bm{u}$
of the kinetic energy~\cite{9780511546099}.
In this work,
the vertical kinetic energy
 \begin{align}
  K_{z\bm{k}} = K_{\| \bm{k}}
  = \frac{K(k)}{8\pi k^2} \left(\frac{k_{\perp}}{k^2}\right)^2
  + \frac{1}{2} \left(K_{\bm{k}} - \frac{K(k)}{4\pi k^2}\right) \left(\frac{k_{\perp}}{k^2}\right)^2
  + \frac{1}{2} (K_{\mathrm{w}\bm{k}}-K_{\mathrm{v}\bm{k}}) \left(\frac{k_{\perp}}{k^2}\right)^2
,
  \label{eq:Kzk}
 \end{align}
 and its polarization anisotropic part, $K_{z \mathrm{PA}\bm{k}}$,
 which is the last term in the right-hand side of Eq.~\eqref{eq:Kzk},
 are used to quantify the anisotropy of a wave-number mode.

\section{Numerical results}
\label{sec:numericalresults}

\subsection{Energy spectra}
\label{ssec:energyspectra}

\begin{figure}[t]
 \begin{center}
  \includegraphics[scale=1]{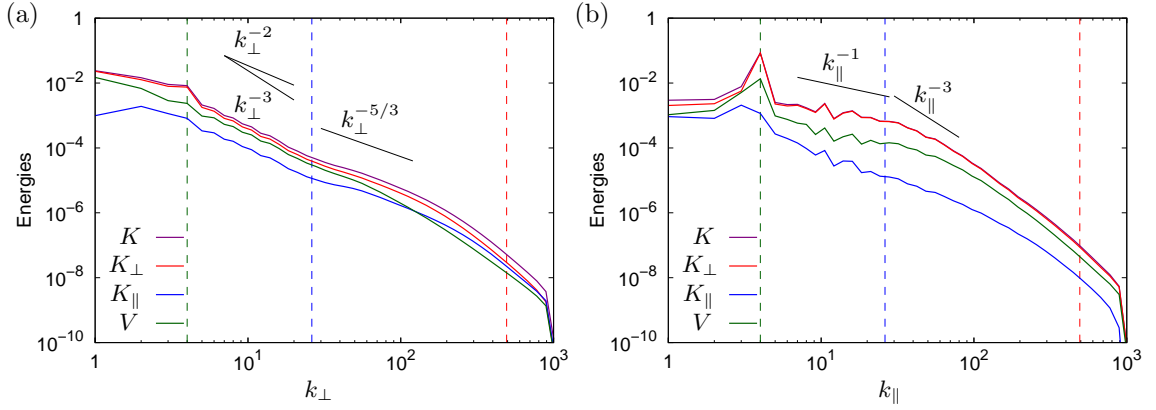}
  \caption{
  Integrated energy spectra:
  total kinetic energy, horizontal kinetic energy,
  vertical kinetic energy, and potential energy.
  (a) as functions of horizontal wave numbers
  integrated over the vertical wave numbers,
  and
  (b) as functions of vertical wave numbers
  integrated over the horizontal wave numbers.
  The green, blue and red vertical dashed lines respectively show
  the forced wave number $k_{\mathrm{f}}$,
  the buoyancy wave number $k_{\mathrm{b}}$, and
  the Ozmidov wave number $k_{\mathrm{O}}$.
  }
  \label{fig:spobs}
 \end{center}
\end{figure}

Spectra of
total kinetic energy $K$, horizontal kinetic energy $K_{\perp}$,
vertical kinetic energy $K_{\|}$, and potential energy $V$
obtained in the numerical simulations with $2048^3$ grid points
are shown in Fig.~\ref{fig:spobs}.
The one-dimensional total kinetic-energy spectrum as a function of the horizontal wave numbers,
for example,
is defined as
\begin{align}
 K(k_{\perp}) =
\frac{1}{\Delta k_{\perp}}
 {\sum_{\bm{k}_{\perp}^{\prime}}}^{\prime}
\sum_{k_{\|}^{\prime}}
 \frac{1}{2} \langle |\bm{u}_{\bm{k}_{\perp}^{\prime}, k_{\|}^{\prime}}|^2\rangle
 ,
\end{align}%
where the summation $\sum_{\bm{k}_{\perp}^{\prime}}^{\prime}$
is taken over $||\bm{k}_{\perp}^{\prime}| - k_{\perp}| < \Delta k_{\perp}/2$,
and $\Delta k_{\perp}$ is the bin width to obtain the spectrum.
The summation $\sum_{k_{\|}^{\prime}}$ is taken over all the vertical wave number.
These one-dimensional spectra are referred to as integrated spectra in this paper.
Figure~\ref{fig:spobs}(a) shows the energy spectra as functions of the horizontal wave numbers
integrated over the vertical wave numbers,
while
Fig.~\ref{fig:spobs}(b) shows those as functions of the vertical wave numbers
integrated over the horizontal wave numbers.
Note that
although the forced wave number $k_{\mathrm{f}}$ is marked for reference in the figures,
the forced wave numbers exist
in the range $k_{\perp} < k_{\mathrm{f}}$ and $k_{\|} < k_{\mathrm{f}}$
because $|\bm{k}|=(k_{\perp}^2 + k_{\|}^2)^{1/2}$.
The buoyancy wave number $k_{\mathrm{b}}$ and
the Ozmidov wave number $k_{\mathrm{O}}$
have the same property.
The integrated spectra show that
the kinetic energy comes mostly from the horizontal component,
and
the potential energy spectra lies between the horizontal and vertical kinetic-energy spectra
for all the wave numbers except for the horizontal wave-number spectra at the very large horizontal wave numbers.

The horizontal wave-number spectra have
a relatively steep spectrum close to $k_{\perp}^{-2}$
at the small horizontal wave numbers,
and a less steep spectrum that is approximately $k_{\perp}^{-5/3}$ at the large horizontal wave numbers.
The transition is observed approximately at the buoyancy wave number
as reported in Ref.~\cite{doi:10.1063/1.3599699}.
However, the energy spectrum at the large $k_{\perp}$ in Fig.~\ref{fig:spobs}(a)
is much steeper than that reported in Ref.~\cite{doi:10.1063/1.3599699},
where the Kelvin-Helmholtz billows are supposed to generate the bump at the large horizontal wave numbers.
It is worth pointing out that
the computational box is flatter
and that the hyper viscosity and the hyper diffusion are used
in the simulation in Ref.~\cite{doi:10.1063/1.3599699}.
Because of the flat computational box,
the bump consists of the large vertical wave-number modes.
The less steep energy spectra appear
near the dissipation range in the inertial subrange,
and they are due to the so-called bottleneck effect.
The hyper viscosity and the hyper diffusion are known
to enhance the bottleneck effect.
One may observe that
this horizontal wave-number spectrum is proportional to $k_{\perp}^{-5/3}$
in all the inertial subrange without any transition,
but there actually exists a transition as seen below.
Similar transition was observed in Refs.~\cite{brethouwer_billant_lindborg_chomaz_2007,FLM:8539071}.
Note that the range of the 3D Kolmogorov turbulence is too small to observe in the spectrum
because the buoyancy Reynolds number
 is evaluated approximately as $2.1$.
The vertical wave-number spectra are also non-uniform,
and the power laws at the small wave numbers and the large wave numbers
 are respectively close to those in Ref.~\cite{GM_ARF} and Ref.~\cite{doi:10.1175/1520-0469(1987)044<1404:EFASSO>2.0.CO;2}.
Similarly to the horizontal wave-number spectra,
the gradual transition is observed roughly at the buoyancy wave number.
The steep spectra similar to $k_{\|}^{-3}$ in the range $k_{\mathrm{b}} < k_{\|} < k_{\mathrm{O}}$
are due to balance between the inertia and the buoyancy~\cite{:/content/aip/journal/pof2/13/6/10.1063/1.1369125,PhysRevFluids.2.104802}.
It is evident in these integrated spectra
that the energy distribution is not scale-invariant
and the energy spectra in the 2D domain spanned by the horizontal and vertical wave numbers show the anisotropy.
It must be emphasized that
these power laws of the integrated spectra consisting of the various slopes
do not necessarily reflect the spectral structures unaffected by the boundary conditions.
In this paper,
the anisotropic energy distribution will be directly investigated below.

\begin{figure}[t]
 \begin{center}
  \includegraphics[scale=.9]{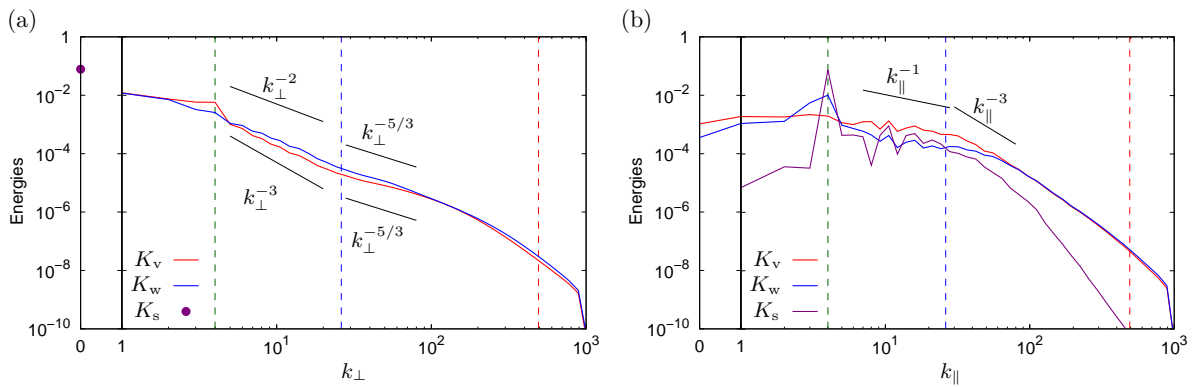}
  \caption{
  (a) Horizontal wave-number spectra
  and (b) vertical wave-number spectra
  of vortical kinetic energy, wave kinetic energy and shear energy.
  The abscissa is scaled
  linearly for $k_{\perp}, k_{\|} \leq 1$
  and
  logarithmically for $k_{\perp}, k_{\|} \geq 1$.
  See also the caption of Fig.~\ref{fig:spobs} for the vertical lines.
  }
  \label{fig:spkhkv}
 \end{center}
\end{figure}

The coexistence of the different power-law exponents in the energy spectra,
where the transition is observed approximately at the buoyancy wave number,
is also observed in the horizontal wave-number spectra
of the vortical kinetic energy and the wave kinetic energy~(Fig.~\ref{fig:spkhkv}(a)).
While the vortical energy spectrum and the wave kinetic energy spectrum are respectively close to $k_{\perp}^{-3}$ and $k_{\perp}^{-2}$ at the small horizontal wave
numbers,
both energy spectra approximately have $k_{\perp}^{-5/3}$ at the large horizontal wave numbers.
The vertical wave-number spectra in Fig.~\ref{fig:spkhkv}(b) also exhibit the coexistence;
the rather flat spectra appears at the small vertical wave numbers,
and the steep spectra similar to the saturation spectrum $k_{\|}^{-3}$ does at the large vertical wave numbers.
These energy spectra are similar to the ones in Ref.~\cite{FLM:8539071}.
The shear energy is defined only for $k_{\perp} = 0$,
but it is large.
In fact,
the kinetic energies of the vortical, wave, and shear flows
integrated over all the wave numbers
are roughly
$4 \times 10^{-2}$, $4 \times 10^{-2}$ and $8 \times 10^{-2}$,
respectively.
The largest energy appears at $k_{\perp}=0$ and $k_{\|}=4$,
which can be directly excited by the external force,
as the shear energy.
Note that although the external force excites both waves and vortices
as well as the shear flows
at a wave-number mode,
and their amplitudes depend on the wave-number mode
as recognized from the energies at the forced wave numbers
in Fig.~\ref{fig:spkhkv}.

\begin{figure}[t]
 \begin{center}
  \includegraphics[scale=1]{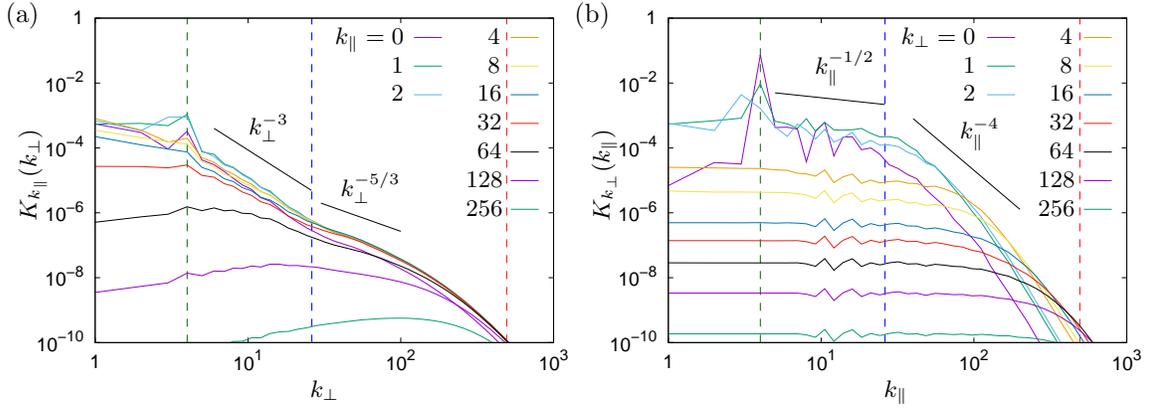}
  \caption{
  Kinetic energy spectra
  (a) for each $k_{\|}$ as function of $k_{\perp}$
  and
  (b) for each $k_{\perp}$ as function of $k_{\|}$.
  See also the caption of Fig.~\ref{fig:spobs} for the vertical lines.
  }
  \label{fig:spslice}
 \end{center}
\end{figure}

The non-uniformity of the horizontal wave-number spectra of the energies shown in Fig.~\ref{fig:spobs}(a)
indicates the existence of the inner structure in the vertical wave-number spectra drawn in Fig.~\ref{fig:spobs}(b)
and vice versa.
The same applies to the vortical kinetic energy and the wave kinetic energy in Fig.~\ref{fig:spkhkv}.
The horizontal wave-number spectra of the energies shown in Fig.~\ref{fig:spobs}(a)
are obtained by integration over the vertical wave numbers,
and the energy spectra without the integration are required to observe the inner structure.
Such non-integrated kinetic-energy spectrum for each $k_{\|}$ as a function of $k_{\perp}$
is defined as
\begin{align}
 K_{k_{\|}}(k_{\perp}) =
\frac{1}{\Delta k_{\perp}}
 {\sum_{\bm{k}_{\perp}^{\prime}}}^{\prime}
\frac{1}{\Delta k_{\|}}
 {\sum_{k_{\|}^{\prime}}}^{\prime}
 \frac{1}{2} \langle |\bm{u}_{\bm{k}_{\perp}^{\prime}, k_{\|}^{\prime}}|^2\rangle
.
\end{align}
The non-integrated kinetic-energy spectrum for each $k_{\perp}$ as a function of $k_{\|}$
is similarly defined.

The non-integrated kinetic-energy spectra are drawn in Fig.~\ref{fig:spslice}.
The kinetic-energy spectra as functions of $k_{\perp}$ for $k_{\|} \leq 32$ shown in Fig.~\ref{fig:spslice}(a)
are not so different from each other,
since the vertical-energy spectra are the rather flat spectra as $k_{\|}^{-1}$ as shown in Fig.~\ref{fig:spobs}(b).
Nevertheless,
we can observe that the energy spectra at small horizontal wave numbers become less steep
roughly from $k_{\perp}^{-3}$ to $k_{\perp}^{-2}$.
As $k_{\|}$ increases further,
the maximal wave number moves to larger $k_{\perp}$.
Most of the kinetic energy at small $k_{\|}$ exists in $k_{\perp} \leq 2$ as shown in Fig.~\ref{fig:spslice}(b).
The integrated energy spectra as functions of the vertical wave numbers shown in Fig.~\ref{fig:spobs}(b)
consist of the corresponding non-integrated energy spectra in $k_{\perp} \leq 2$.
It is consistent with the fact that the horizontal wave-number spectra uniformly and rapidly decrease
as shown in Fig.~\ref{fig:spobs}(a).
Moreover,
in the range $30 \lessapprox k_{\|} \lessapprox 500$,
the relatively flat spectrum close to $k_{\|}^{-1/2}$ extends to the large $k_{\|}$
as $k_{\perp}$ increases.
Then,
the large $k_{\perp}$ has larger energy at the large $k_{\|}$
than the small $k_{\perp}$ has~\cite{PhysRevFluids.2.104802}.
Thus, the integration over $k_{\perp}$ makes the saturation spectrum complex in the large $k_{\|}$ range.
The saturation spectrum is considered to consist of the breaking of the internal gravity waves~\cite{doi:10.1029/JD091iD02p02742}.
Since the integrated spectra of the kinetic energy shown in Figs.~\ref{fig:spobs}(a) and \ref{fig:spobs}(b) are
respectively obtained by summation of the non-integrated spectra shown in Figs.~\ref{fig:spslice}(a) and \ref{fig:spslice}(b),
the integrated spectra are determined mostly by the non-integrated spectra in the few small codimensional wave numbers.
In this sense,
the integrated spectra cannot properly reflect the energy distribution at the moderate wave numbers
unaffected by the boundary conditions.
Moreover,
the identification of the dominant physical mechanism by the integrated spectra
requires a careful inspection.

\begin{figure}[t]
 \begin{center}
  \includegraphics[scale=.9,clip]{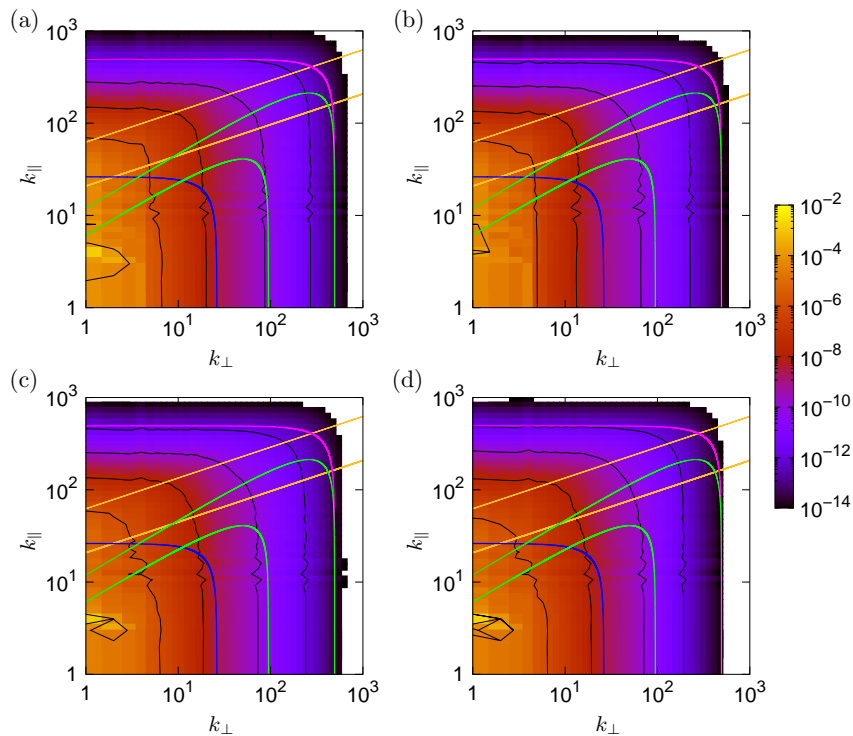}
  \caption{
  2D spectra of
  (a) total kinetic energy,
  (b) vortical energy,
  (c) wave kinetic energy,
  and
  (d) potential energy.
  The contours are drawn for $10^{-12}, 10^{-10}, \cdots, 10^{-4}$.
  The critical wave number at which $\chi_{\bm{k}}=1/3$
  and that at which $\chi_{\bm{k}}=1$
  are represented by the thick and thin green curves,
  respectively.
  The 2D critical wave number at which $\chi_{\mathrm{2D}\bm{k}}=1/3$
  and that at which $\chi_{\mathrm{2D}\bm{k}}=1$
  are represented by the thick and thin yellow curves,
  respectively.
  The buoyancy wave number and Ozmidov wave number are respectively represented by the blue and magenta curves.
  }
  \label{fig:spKvKwmap}
 \end{center}
\end{figure}

To observe the anisotropic structures of the energy spectra,
the 2D spectra for total, vortical, wave kinetic, and potential energies
in the horizontal and vertical wave-number domain
are drawn in Fig.~\ref{fig:spKvKwmap},
which provides an overview of the energy spectra.
The 2D spectrum is defined as
\begin{align}
 K(k_{\perp}, k_{\|}) =
\frac{1}{2\pi k_{\perp}}
\frac{1}{\Delta k_{\perp}}
 {\sum_{\bm{k}_{\perp}^{\prime}}}^{\prime}
\frac{1}{\Delta k_{\|}}
 {\sum_{k_{\|}^{\prime}}}^{\prime}
 \frac{1}{2} \langle |\bm{u}_{\bm{k}_{\perp}^{\prime}, k_{\|}^{\prime}}|^2\rangle
,
\end{align}
where the normalizing constant, $1/ (2\pi k_{\perp})$,
is introduced
for the contours of the energy spectra
to be compared easily with the completely isotropic ones.

All the energies shown in Fig.~\ref{fig:spKvKwmap}
 accumulate at small $k_{\perp}$.
 It is consistent with the large energies at small $k_{\perp}$
 in the integrated and non-integrated spectra
 shown in Figs.~\ref{fig:spobs}--\ref{fig:spslice}.
 The energies drawn as the 2D spectra
 obviously show the anisotropy in small $k_{\perp}$ and $k_{\|}$.
As $k = \sqrt{k_{\perp}^2+k_{\|}^2}$ becomes large,
 the contours of each energy
 are more similar to the isotropic curves
 which show the buoyancy wave number and the Ozmidov wave number.
 Such fact indicates that the anisotropy that exists at the small $k$
 gradually decreases
 and the flow at these scales is closer to the 3D isotropic Kolmogorov turbulence,
 as $k$ become large.
 Note that even at the Ozmidov wave number
 the energy in $k_{\perp} < k_{\|}$ is larger than that in $k_{\perp} > k_{\|}$,
 and the energy spectra are still weakly anisotropic.

 It is not clear
 in Fig.~\ref{fig:spKvKwmap}
 where the wave kinetic energy and the potential energy are larger than the vortical energy.
  Furthermore,
  the four 2D energy spectra may appear close enough.
 However,
 by careful observation,
 we can find that
 the spectra of the wave kinetic energy (Fig.~\ref{fig:spKvKwmap}(c)) and the potential energy (Fig.~\ref{fig:spKvKwmap}(d))
 are similar,
 but the vortical-energy spectrum (Fig.~\ref{fig:spKvKwmap}(b)) is different from these.

\subsection{Distribution of turbulence indices in wave-number space}
\label{ssec:cbrange}

It is indispensable to separate the wave-number space
based on the dominant physical mechanisms of turbulence.
In particular,
the theory of the critical balance needs the separation of the wave-dominant range.
To quantitatively discuss
whether the balance between linear and nonlinear time scales can identify the wave-dominant range,
the energy decomposition written in Sec.~\ref{ssec:decomposition} is employed for the definition.

\begin{figure}[t]
 \begin{center}
\includegraphics[scale=.9]{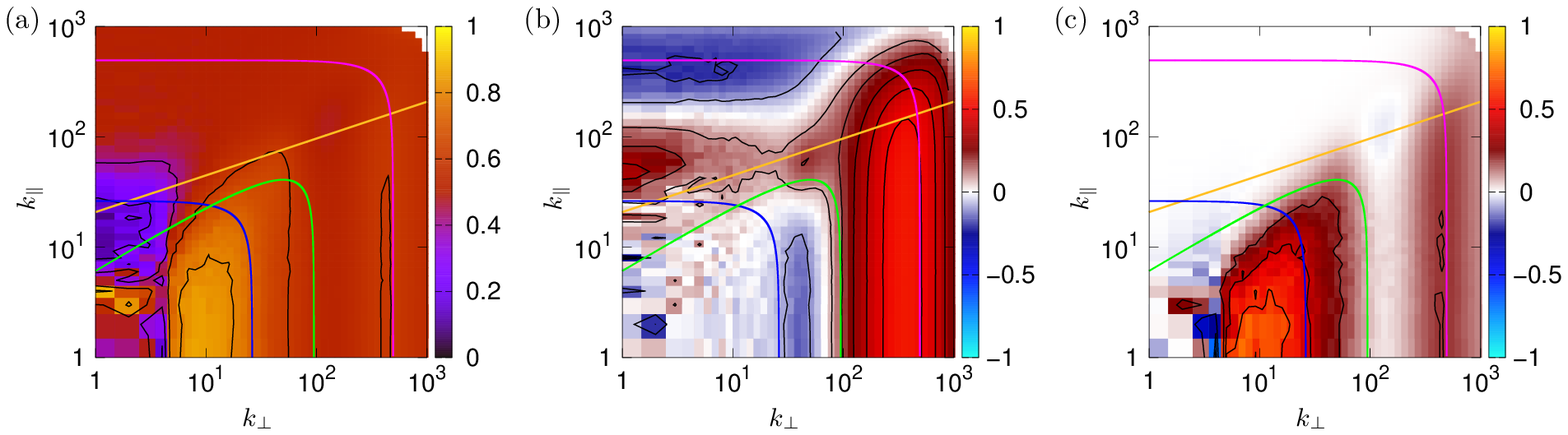}
  \caption{
  (a) ratio of the wave kinetic energy to the total kinetic energy
  $K_{\mathrm{w}\bm{k}}/K_{\bm{k}}$,
  (b) relative difference between the wave kinetic energy and the potential energy
   $(K_{\mathrm{w}\bm{k}} - V_{\bm{k}})/(K_{\mathrm{w}\bm{k}} + V_{\bm{k}})$,
  and
  (c) ratio of the polarization anisotropic part to the kinetic energy $K_{z \mathrm{PA}\bm{k}}/K_{\bm{k}}$.
  The contours are drawn for every $0.2$ in (a) and (c),
  and for every $0.1$ in (b).
  The critical wave number at which $\chi_{\bm{k}}=1/3$,
  the 2D critical wave number at which $\chi_{\mathrm{2D}\bm{k}}=1/3$,
  the buoyancy wave number, and the Ozmidov wave number
  are represented by the green, yellow, blue, and magenta curves,
  respectively.
  }
  \label{fig:Kpa_Kz}
 \end{center}
\end{figure}

The difference of the vortical energy
 from the wave kinetic energy and the potential energy,
 and the similarity of the wave kinetic energy and the potential energy
 can be used to characterize the wave turbulence and the strong turbulence.
 In the wave-dominant range,
 the wave kinetic energy is postulated to be much larger than the vortical energy.
 The weak nonlinearity assumes that
 the wave kinetic energy is also expected to be close to the potential energy in the same range.
 Since the energies are not uniform in the wave-number space,
 a normalization of the energy is required to characterize each range;
 the ratios of the energies are drawn in Fig.~\ref{fig:Kpa_Kz}
  to quantify the dominance of the weak-wave turbulence.
 For example,
 the ratio of the wave kinetic energy to the total kinetic energy
 is used
 instead of direct comparison between the wave kinetic energy and the vortical energy.

 The ratio of the wave kinetic energy to the total kinetic energy
 \begin{align}
  \frac{K_{\mathrm{w}\bm{k}}}{K_{\bm{k}}}
  = \frac{K_{\mathrm{w}\bm{k}}}{K_{\mathrm{v}\bm{k}}+K_{\mathrm{w}\bm{k}}+K_{\mathrm{s}\bm{k}}}
\label{eq:wave2total}
 \end{align}
 is drawn in Fig.~\ref{fig:Kpa_Kz}(a).
Note that the shear kinetic energy is defined only on $k_{\perp}=0$,
and it does not appear in Fig.~\ref{fig:Kpa_Kz}(a).
The weak turbulence theory requires that
the linear time scale is much shorter than the nonlinear time scale,
and the ratios of the nonlinear time scale to the linear time scale $\chi_{\bm{k}}$
are usually $O(0.1)$.
See Ref.~\cite{PhysRevE.89.012909} for example.
It was reported in magnetohydrodynamic turbulence that
the wave numbers
at which the ratio of the nonlinear time scale to the linear time scale $\chi_{\bm{k}}=1/3$
are the critical wave numbers
separating the weak and strong turbulence~\cite{PhysRevLett.116.105002}.
Note that
the value $1/3$ is introduced as a rough indication
because the transition between the wave-dominant range and the vortex-dominant range
is gradual.
In the present numerical simulation,
the contour of $K_{\mathrm{w}\bm{k}}/K_{\bm{k}}=0.6$ is close to the curve of $\chi_{\bm{k}} = 1/3$.
The wave kinetic energy is dominant in the total kinetic energy
over the vortical energy
at the wave numbers where $\chi_{\bm{k}} \lessapprox 1/3$.
Note that the range of $k_{\perp}, k_{\|} < 5$ is directly affected by the external force,
and is not considered here.

The dominance of the wave-kinetic energy
does not always results in the weak-wave turbulence~\cite{kafiabad_bartello_2018}.
In the weak-wave turbulence,
the wave-number modes must have the wave kinetic energy close to the potential energy.
The relative difference between the wave kinetic energy and the potential energy
 \begin{align}
  \frac{K_{\mathrm{w}\bm{k}} - V_{\bm{k}}}{K_{\mathrm{w}\bm{k}} + V_{\bm{k}}}
  \label{eq:weaknonlinear}
 \end{align}
is drawn in Fig.~\ref{fig:Kpa_Kz}(b).
In the weak-wave turbulence,
$K_{\mathrm{w}} \approx V$, i.e.,
it is anticipated that the relative difference is close to $0$
because of the weak nonlinearity.
In fact,
$-0.2 < (K_{\mathrm{w}\bm{k}} - V_{\bm{k}})/(K_{\mathrm{w}\bm{k}} + V_{\bm{k}}) < 0.1$
in the range where $\chi_{\bm{k}} \lessapprox 1/3$.
Therefore,
the wave-number modes
where the wave-kinetic energy is dominant over the vortical energy
coincide with the modes which have the relative difference between the wave kinetic energy and the potential energy close to $0$.
Namely,
the wave-number modes where $\chi_{\bm{k}} \lessapprox 1/3$
is in the weak-wave turbulence.

Moreover,
in Fig.~\ref{fig:Kpa_Kz}(c),
 the ratio of the polarization anisotropic part to the total kinetic energy
 \begin{align}
  \frac{K_{z \mathrm{PA}\bm{k}}}{K_{\bm{k}}}
  = \frac{K_{\mathrm{w}\bm{k}}-K_{\mathrm{v}\bm{k}}}{2K_{\bm{k}}} \left(\frac{k_{\perp}}{k}\right)^2
  \label{eq:kzpa}
 \end{align}
is drawn.
Here,
$K_{z \mathrm{PA}\bm{k}} = \mathrm{Re} [Z_{\bm{k}} h_{+ z \bm{k}}^2]
= (k_{\perp}/k)^2 (K_{\mathrm{w}\bm{k}}-K_{\mathrm{v}\bm{k}}) / 2$
represents the polarization anisotropic part of the vertical kinetic energy
according to the helical-mode decomposition.
Equation~(\ref{eq:kzpa}) indicates
the direct relation between the anisotropy and the dominance of the wave-kinetic energy over the vortical energy given by Eq.~(\ref{eq:wave2total}).
In fact,
the wave-number modes in the weak-wave turbulence,
where $\chi_{\bm{k}} \lessapprox 1/3$,
has $K_{z \mathrm{PA}\bm{k}}/K_{\bm{k}} > 0.2$.
The weak-wave turbulence of internal gravity waves has strong anisotropy.

The ratio of the gravity-wave period to the eddy-turnover time $\chi_{\bm{k}}$ well separates the weak-wave turbulence
also from the horizontally long waves $k_{\perp} \approx 1$ and $k_{\|} \sim O(10)$.
The 2D ratio $\chi_{\mathrm{2D}\bm{k}}$ also does it
if $\chi_{\mathrm{2D}\bm{k}} = 1/3$ is selected as a threshold,
though $\chi_{\mathrm{2D}\bm{k}}$ cannot separate the weak-wave turbulence from the 3D isotropic Kolmogorov turbulence by definition.

The wave-number range of the anisotropic weak-wave turbulence
is smaller than the inner range of the Ozmidov wave number.
The transient wave-number range
from the anisotropic weak-wave turbulence to the 3D isotropic Kolmogorov turbulence
appears in the middle of the two turbulence range,
where the quasi-2D turbulence is dominant.
In this transient range,
the eddy-turnover time of the wave-number mode is larger than the Brunt-V\"ais\"al\"a period
and is smaller than $1/3$ of the linear wave period of the mode.
i.e., $1/N \lessapprox \tau_{\bm{k}} \lessapprox 3/\sigma_{\bm{k}}$,
and the range is noticeable at the small horizontal and large vertical wave numbers.
The wave-breaking is known to occur mainly at the small horizontal and large vertical wave numbers~\cite{doi:10.1175/1520-0485(1977)007<0836:EMWTOI>2.0.CO;2}.
The saturation spectrum $K(k_{\|}) \propto k_{\|}^{-3}$ is observed in this range
as shown in Fig.~\ref{fig:spobs}(b).

In the wave-number range $k_{\perp} \gg k_{\|}$,
$K_{\mathrm{w}} = (k/ k_{\perp})^2 K_{\|}  \approx K_{\|}$,
and $K_{\mathrm{v}} \approx K_{\perp}$.
Therefore, the horizontal energy spectrum $K_{\perp} \propto k_{\perp}^{-5/3}$
shown in Fig.~\ref{fig:spobs}(a)
results mainly from the vortical mode.
The fact that $K_{\mathrm{w}} > K_{\mathrm{v}}$
indicates that $K_{\|} > K_{\perp}$ in the wave-number range,
which is confirmed by drawing $K_{\|} / K$ though the figure is omitted.
The weak-wave turbulence is stronger than the quasi-2D turbulence
in the range where $\chi_{\bm{k}} \lessapprox 1/3$.
In addition,
the quasi-2D turbulence, i.e., the pancake turbulence~\cite{:/content/aip/journal/pof2/13/6/10.1063/1.1369125}
is dominant in the small $k_{\perp}$ and large $k_{\|}$ range
where $\chi_{\bm{k}} \gtrapprox 1/3$ and $k<k_{\mathrm{O}}$.

\begin{figure}
 \begin{center}
  \includegraphics[scale=.9]{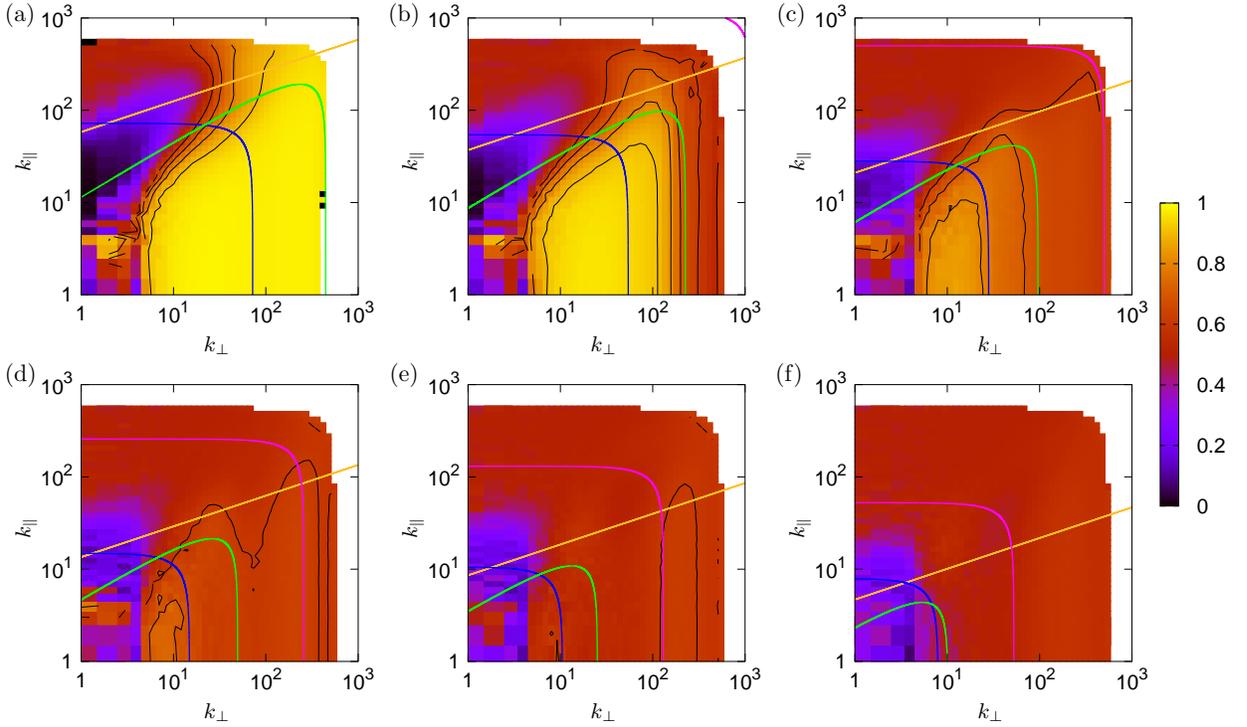}
  \caption{%
  Ratio of the wave kinetic energy to the total kinetic energy
  $K_{\mathrm{w}\bm{k}}/K_{\bm{k}}$.
  (a) $\gamma_{\bm{k}}=0.1$, (b) $0.2$, (c) $0.5$,
  (d) $1$, (e) $2$, and (f) $5$.
  See also the caption of Fig.~\ref{fig:Kpa_Kz} for the curves.
  }
  \label{fig:parameter}
 \end{center}
\end{figure}

The wave period given by the linear dispersion relation
characterizes the weak-wave turbulence
better than the the Brunt-V\"ais\"al\"a period
as seen in Fig.~\ref{fig:Kpa_Kz}.
To confirm it,
the ratios of the wave kinetic energy to the total kinetic energy
for different amplitudes of the external force
are drawn in Fig.~\ref{fig:parameter}.
The numerical simulations to draw Fig.~\ref{fig:parameter} are performed
by using $1024^3$ grid points.
The amplitude of the external force $\gamma_{\bm{k}}$ is varied from $0.1$ to $5$
in the low-resolution simulations
for comparison with $\gamma_{\bm{k}}=0.5$,
which is used to draw Figs.~\ref{fig:spobs}--\ref{fig:Kpa_Kz}.

The range of the weak-wave turbulence is the largest when the external force is the smallest (Fig.~\ref{fig:parameter}(a)),
and the range becomes smaller as the external force is larger. (Figs.~\ref{fig:parameter}(b)--\ref{fig:parameter}(e))
It results from the fact that the eddy-turnover time becomes smaller as the turbulent fluctuation is more excited.
The threshold $\chi_{\bm{k}} = 1/3$ well separates the weak-wave turbulence
independently of the buoyancy Reynolds number and the vertical Froude numbers considered here.
For $\gamma_{\bm{k}} = 1$,
the wave-number range of $\chi_{\bm{k}} \lessapprox 1/3$
and hence the number of the wave-number modes are small.
(Fig.~\ref{fig:parameter}(f))
Then,
the weak-wave turbulence cannot be organized
because the resonant interactions are rare.
Such divergence in the simulation with this large external force is consistent with the break in the monotonicity of the Reynolds number and the vertical Froude number in Table~\ref{tab:parameters}.
It is derived from the limitation of numerical simulations due to the discretization and the periodic boundary condition.
The wave-dominant range should exist even for this buoyancy Reynolds number
and the Froude number,
if the simulations in a much larger computational domain,
which provides denser grid points in the wave-number space,
were performed.

\subsection{Deviation from linear dispersion relation in wave-number space}

It has been exhibited in the previous subsection
that the ratios of the nonlinear time scale to the linear time scale $\chi_{\bm{k}}$
i.e., the characteristic times
can successfully separate the wave-dominant range
by using the Cartesian decomposition and the helical-mode decomposition.
To observe that the dominance of the waves in the range where $\chi_{\bm{k}} \lessapprox 1/3$
in another way,
a frequency deviation from the linear dispersion relation is evaluated.
It is convenient to introduce a complex amplitude
used in the weak turbulence theory~\cite{zak_book}.
The complex amplitude in the present system is defined as
\begin{align}
 a_{\bm{k}} = \frac{1}{\sqrt{2 \sigma_{\bm{k}}}} \left(u_{z\bm{k}} - \frac{i}{N} b_{\bm{k}}\right).
\end{align}
Because the linear inviscid non-diffusive equation~(\ref{eq:linearinviscidnondiffusive}) can be rewritten as
$\partial a_{\bm{k}}/ \partial t = - i \sigma_{\bm{k}} a_{\bm{k}}$,
the frequency spectrum of $a_{\bm{k}}$ has a value
only at $-\sigma_{\bm{k}} = -Nk_{\perp}/k$
in the linear inviscid non-diffusive limit.
The minus sign in front of the frequency comes from the conventional expression
of the canonical equation in the weak turbulence theory.
A frequency deviation is defined as
\begin{align}
\delta \sigma_{\bm{k}}
 = \left( \frac{\displaystyle \sum_{\sigma} (\sigma + \sigma_{\bm{k}})^2 |\widetilde{a}_{\bm{k},\sigma}|^2}{\displaystyle \sum_{\sigma} |\widetilde{a}_{\bm{k},\sigma}|^2} \right)^{\frac{1}{2}}
,
\end{align}
where $\widetilde{a}_{\bm{k},\sigma}$ denotes the Fourier coefficient
obtained from the time series of $a_{\bm{k}}(t)$.
The relative frequency deviation, $\delta \sigma_{\bm{k}} / \sigma_{\bm{k}}$,
is employed for the measure of the wave nature of a wave-number mode
in this paper.
When the weakly nonlinear wave mode is dominant at a wave-number mode,
the frequency spectrum is narrow-band and it has a peak at the frequency given by the linear dispersion relation,
and the relative frequency deviation of the wave-number mode is small.
Conversely,
when the nonlinearity is not weak owing to the vortical mode and/or other wave-number modes,
the frequency spectrum is broad-band or it has peaks away from the linear frequency~\cite{kafiabad_bartello_2018},
and the relative frequency deviation is large.
Note that the nonlinearity changes the frequency spectrum in two ways:
one is the excitation of frequencies which do not satisfy the dispersion relation
due to the nonlinear interactions among wave-number modes,
and the other is the frequency shift due to the small-wave-number flows
such as the Doppler effect.

\begin{figure}
 \begin{center}
  \includegraphics[scale=1]{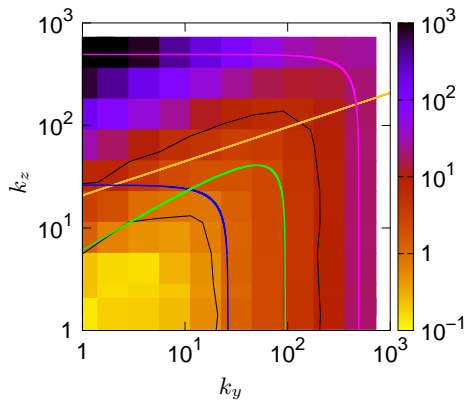}
  \caption{%
  Relative frequency deviation $\delta \sigma_{\bm{k}} / \sigma_{\bm{k}}$
  for $\bm{k} = (k_x,k_y,k_z) = (0, 2^p, 2^q)$ where $p,q=0,1,2,\cdots$.
  The contours are drawn for $1$ and $10$.
  See also the caption of Fig.~\ref{fig:Kpa_Kz} for the curves.
  }
  \label{fig:freqsp}
 \end{center}
\end{figure}

The relative frequency deviation is drawn in Fig.~\ref{fig:freqsp}.
The frequency spectra are obtained
from the time series of $a_{\bm{k}}$,
where $\bm{k} = (k_x,k_y,k_z) = (0, 2^p, 2^q)$ and $p,q=0,1,2,\cdots$,
in the high-resolution simulation.
The relative frequency deviation is small
in the range where $\chi_{\bm{k}} \lessapprox 1/3$,
and shows similarity to $\chi$,
becoming large as $\chi$ increases.
This results from the increase of the band width of the frequency spectrum
due to the nonlinearity.
One may notice that
the difference
between the contours of the relative frequency deviation and $\chi$
at the large horizontal wave numbers where $k_y=32, 64$ and $k_z\leq 16$
is relatively large.
The difference can be interpreted by the Doppler shift due to
the horizontal flows with the small horizontal wave numbers
including the vertically-sheared horizontal flows
having most of the total energy in the flow field
as recognized from Figs.~\ref{fig:spobs}--\ref{fig:spslice}.
Then, the dominance of the weakly nonlinear wave mode
in the range where $\chi_{\bm{k}} \lessapprox 1/3$
is supported by the frequency deviation of wave-number modes.

\section{Concluding remark}
\label{sec:summary}

In this paper,
direct numerical simulations of strongly stratified turbulence
where the internal gravity-wave turbulence and the strong turbulence coexist
were performed.
The energies accumulate at the small horizontal wave numbers,
and the energies at the small vertical wave numbers are also large.
Then,
the one-dimensional spectra,
which are obtained by the integration over the horizontal or vertical wave numbers,
or by using the norm of the wave-number vector,
mask the inner structures,
and do not appropriately represent the critical wave numbers
separating the wave-number range of the weak-wave turbulence.
The non-integrated spectra and the two-dimensional spectra drawn in the horizontal and vertical wave-number domain
reveal the inner structures of the anisotropic turbulence.
The results show that the power laws observed in the one-dimensional spectra
are superposition of various distributions of the spectral amplitude.
Therefore,
much care should be taken when the spectra are compared with the experimentally observed spectra,
which are mostly obtained from one-dimensional time series.

Following the premise that
the wave kinetic energy is much larger than the vortical energy,
and is close to the potential energy in the range of the weak-wave turbulence,
non-dimensional indices
based on the energies,
Eqs.~(\ref{eq:wave2total})--(\ref{eq:weaknonlinear}),
were proposed to determine the range in the wave-number space.
It was also clarified
by another non-dimensional index based on the energies, Eq.~(\ref{eq:kzpa}),
that
the polarization anisotropy in the range is large,
resulting from the wave kinetic energy being larger than the vortical energy.
These non-dimensional indices proposed in this paper show the similar distribution,
which confirms the appropriateness of the indices for the identification of the range of the anisotropic weak-wave turbulence.
The dominance of the waves in the range is also verified
by the frequency spectra having peaks at the frequency given by the linear dispersion relation.

From the distributions of the non-dimensional indices in the horizontal and vertical wave-number domain,
it was found that
the range,
which emerges at the small horizontal and vertical wave numbers,
is anisotropic
and smaller than the inner range of the Ozmidov wave number.
The wave-number modes in the weak-wave turbulence
have the linear time scale given by the linear dispersion relation
smaller than $1/3$ of the nonlinear time scale, which is the eddy-turnover time.
In other words,
the critical wave number which separates the weak-wave turbulence
has the ratio of the linear time scale to the nonlinear one being $1/3$.
In most anisotropic turbulence systems,
we have some options for linear and nonlinear time scales.
The present results show that
the range of the anisotropic weak-wave turbulence in the wave-number space can be identified
when the appropriate time scales are selected
in consideration of the anisotropy of the flow field.

 The difference between the linear period given by the linear dispersion relation
 and the isotropic Brunt-V\"ais\"al\"a period
 is large in the range where the horizontal wave numbers are small and the vertical wave numbers are relatively large.
 The dynamics in the wave-number range is determined
 neither by the weak-wave turbulence nor by the three-dimensional isotropic Kolmogorov turbulence.
 The wave breaking is dominant in this wave-number range~\cite{mccomas-1981-11},
 and it is consistent with the saturation spectrum in Fig.~\ref{fig:spobs}(b).
 The critical balance states the energy transfer from the waves to the eddies in this range.
In this sense,
the coexistence of the waves and eddies might play an important role
in the energy spectrum~\cite{waite_bartello_2004,waite_bartello_2006}.
 The critical balance is the energy transfer
 in such transitional wave-number range between the wave-dominant and vortex-dominant ranges.
 The separation of the weak-wave turbulence in the present paper suggests that
 the critical balance should appear in the wave-number range $1/N \lessapprox \tau_{\bm{k}} \lessapprox 3/\sigma_{\bm{k}}$.
 The energy transfer in the transitional wave-number range will be reported elsewhere.

\begin{acknowledgments}
Numerical computation in this work was carried out
 at the Yukawa Institute Computer Facility, Kyoto University
 and Research Institute for Information Technology, Kyushu University.
This work was partially supported by JSPS KAKENHI Grant
No.~15K17971, No.~16K05490, No.~17H02860, and No.~18K03927.
\end{acknowledgments}

\end{document}